\documentclass[conference,a4paper]{IEEEtran}
\usepackage[left=1in, right=0.7in, bottom=1.4in, top=0.76in]{geometry}
\IEEEoverridecommandlockouts

\usepackage{cite}
\usepackage{amsmath,amssymb,amsfonts}
\usepackage{algorithmic}
\usepackage{graphicx}
\usepackage{textcomp}
\usepackage{xcolor}
\usepackage{fancyhdr}
\def\BibTeX{{\rm B\kern-.05em{\sc i\kern-.025em b}\kern-.08em
    T\kern-.1667em\lower.7ex\hbox{E}\kern-.125emX}}

\usepackage{array}
\usepackage{acronym}

\acrodef{FEEC}{Faculty of Electrical Engineering and Communication}
\acrodef{BUT}{Brno University of Technology}
\acrodef{CIR}{channel impulse response}
\acrodef{RMS}{root mean square}
\acrodef{DS}{delay spread}
\acrodef{MMW}[MMW]{millimeter wave}
\acrodef{FMCW}[FMCW]{frequency modulated continuous wave}
\acrodef{FSPL}[FSPL]{free space path loss}
\acrodef{FFT}[FFT]{fast fourier transform}
\acrodef{IFFT}[IFFT]{inverse fast fourier transform}
\acrodef{USB}[USB]{upper sideband}
\acrodef{LSB}[LSB]{lower sideband}
\acrodef{IRR}{image rejection ratio}
\acrodef{SNR}{signal to noise ratio}
\acrodef{MPC}{multipath components}
\acrodef{LOS}{line-of-sight}
\acrodef{NLOS}{non-line-of-sight}
\acrodef{GPS}{global positioning system}
\acrodef{TX}{transmitter}
\acrodef{RX}{receiver}
\acrodef{ATT}{attenuator}
\acrodef{G}{gain}
\acrodef{RP}{relative power}
\acrodef{PDP}{power delay profile}
\acrodef{OWGA}{open waveguide antenna}
\acrodef{HPBW}{half power beam width}
\acrodef{AGV}{autonomous ground vehicles}
\acrodef{CDF}{cumulative distribution function}
\acrodef{WRC}{world radiocommunication conference}

\begin{document}
\makeatletter

\title{Long-Term Channel Analysis at \mbox{60 and 80 GHz} for \mbox{Autonomous Ground Vehicles}}

\author{\IEEEauthorblockN{ Radek Zavorka\IEEEauthorrefmark{1}, Tomas Mikulasek\IEEEauthorrefmark{1}, Josef Vychodil \IEEEauthorrefmark{1}, Jiri Blumenstein\IEEEauthorrefmark{1}, Hussein Hammoud\IEEEauthorrefmark{2},\\ Wojtuń Jarosław \IEEEauthorrefmark{4}, Aniruddha Chandra \IEEEauthorrefmark{3}, Jan M. Kelner \IEEEauthorrefmark{4}, Cezary Ziółkowski \IEEEauthorrefmark{4}, Ales Prokes \IEEEauthorrefmark{1}}
\IEEEauthorblockA{\IEEEauthorrefmark{1}Department of Radio Electronics, Brno University of Technology, Brno, Czech Republic}
\IEEEauthorblockA{\IEEEauthorrefmark{2}University of Southern California, Los Angeles, USA}
\IEEEauthorblockA{\IEEEauthorrefmark{3}National Institute of Technology, Durgapur, India}
\IEEEauthorblockA{\IEEEauthorrefmark{4}
Institute of Communications Systems, Faculty of Electronics,\\ Military University of Technology, 00-908 Warsaw, Poland}
e-mail: xzavor03@vutbr.cz}

\fancypagestyle{firstpage}
{
    \fancyhead[C]{The paper has been presented at the 2024 IEEE Conference on Antenna Measurements and Applications (CAMA), Da Nang, Vietnam, October 9-11, 2024}    
    \fancyfoot[C]{This research was funded in part by the National Science Center (NCN), Poland, grant no. 2021/43/I/ST7/03294 (MubaMilWave). For this purpose of Open Access, the author has applied a CC-BY public copyright license to any Author Accepted Manuscript (AAM) version arising from this submission.}
}

\maketitle

\thispagestyle{firstpage}

\begin{abstract}
This paper presents a comprehensive measurement campaign aimed at evaluating indoor-to-indoor radio channels in dynamic scenarios, with a particular focus on applications such as autonomous ground vehicles (AGV). These scenarios are characterized by the height of the antennas, addressing the unique challenges of near-ground communication. Our study involves long-term measurements (20 minutes of continuous recording per measurement) of the channel impulse response (CIR) in the 60\,GHz and 80\,GHz frequency bands, each with a bandwidth of 2.048\,GHz. We investigate the variations in channel characteristics, focusing on parameters such as root mean square (RMS) delay spread and the \mbox{Rician $K$-factor.}

\end{abstract}


\section{Introduction}
\label{introduction}
In wireless communication, understanding how radio channels behave in dynamic environments is key to ensuring reliable performance. This paper focuses on a comprehensive measurement campaign aimed at evaluating indoor-to-indoor radio channels in dynamic scenarios. Specifically, our study is designed with emphasis on applications such as \ac{AGV}, where changing conditions can significantly affect communication quality.

Two promising frequency bands for high-speed data transmission were analyzed, 60\,Ghz and 80\,Ghz, respectively. The 60\,Ghz band, which offers several GHz of bandwidth, has been assigned by the International Telecommunication Union (ITU) to the Industrial, Scientific, and Medical (ISM) bands, enabling license-free operations~\cite{ITU_ISM_60}. Additionally, the E-band covers the 71-76\,Ghz and 81-86\,Ghz bands, offering 5\,Ghz of bandwidth, and has been considered as a candidate for 5G at \mbox{\ac{WRC}-15} \cite{wrc19}. Compared to the 60\,Ghz band, E-band waves are less prone to oxygen absorption during atmospheric propagation. \cite{absorb}.

Dynamic scenarios introduce unique challenges to wireless communication systems. In indoor environments, movement of people, furniture, and other objects can alter signal propagation paths, leading to variability in signal strength and multipath effects \cite{mpc}. These fluctuations can impact communication reliability, latency, and throughput, which are critical for applications that rely on stable wireless connections \cite{industry}.

Our measurement campaign involved capturing extensive data across a variety of indoor settings, focusing on how the radio channel changes over time. By analyzing this data, we aim to identify statistical patterns and evaluate their impact on communication performance. The analysis includes examining metrics such as \ac{RMS} delay spread and Rician $K$-factor, each of which offers insights into the nature of the radio channel and its potential influence on communication systems \cite{MPC_nemecko}.

Maintaining regular communication is essential for the effectiveness and safety of \ac{AGV} operating indoors \cite{agv1}. In applications like warehouse automation and smart transportation, \ac{AGV} play a pivotal role \cite{agv1, agv2}. For navigation, decision-making, and interacting with other systems, they depend on reliable wireless connection. However, operating in indoor environments introduces challenges due to changing layouts, moving obstacles, and varied signal paths, which can cause significant fluctuations in strength of radio signals. The height of communication antennae, a challenge known as near-ground propagation \cite{AGV_molisch, AGV_IEEE1, AGV_IEEE2, AGV_MDPI}, is specified for \ac{AGV}.




\subsection{Contribution of the Paper}
In this paper, we present our measurements of the time-variant channel in the frequency band of 60\,GHz and 80\,GHz with a bandwidth of 2.048\,GHz. Our objective was to investigate the variations in channel characteristics and their parameters such as \ac{RMS} delay spread or Rician $K$-factor.

The main contributions of this paper are as follows:
\begin{itemize}
    \item We present a long-term measurement campaign (20~minutes of continuous recording per measurement) of a dynamic channel in the frequency band of 60\,GHz and 80\,GHz in an indoor-to-indoor scenario with antennas position to simulate differences between \ac{AGV} and pedestrian.
    \item We analyze the impact of a mooving people and objects in the vicinity of \ac{TX} and \ac{RX} antennas and its influence on radio channel properties.
\end{itemize}

\subsection{Paper organization}
The organization of the remaining sections is as follows: Section \ref{section_MCD} provides a detailed description of the measurement campaign. Section \ref{section_MS} discusses the measurement setup, including the equipment and configurations used, with Subsection \ref{section_MS_LTM} explaining the principle of long-term data recording. In Section \ref{section_statistic}, we provide a statistical analysis of the indoor radio channel, focusing on the RMS delay spread and Rician $K$-factor in Subsections \ref{section_statistic_RMS} and \ref{section_statistic_Kfac}, respectively. Finally, in Section \ref{section_C}, we summarize the key findings and present the conclusions of this work.

\section{Measurement campaign description} \label{section_MCD}
\begin{figure}[htbp]
    \centering
    \includegraphics[width=0.48\textwidth]{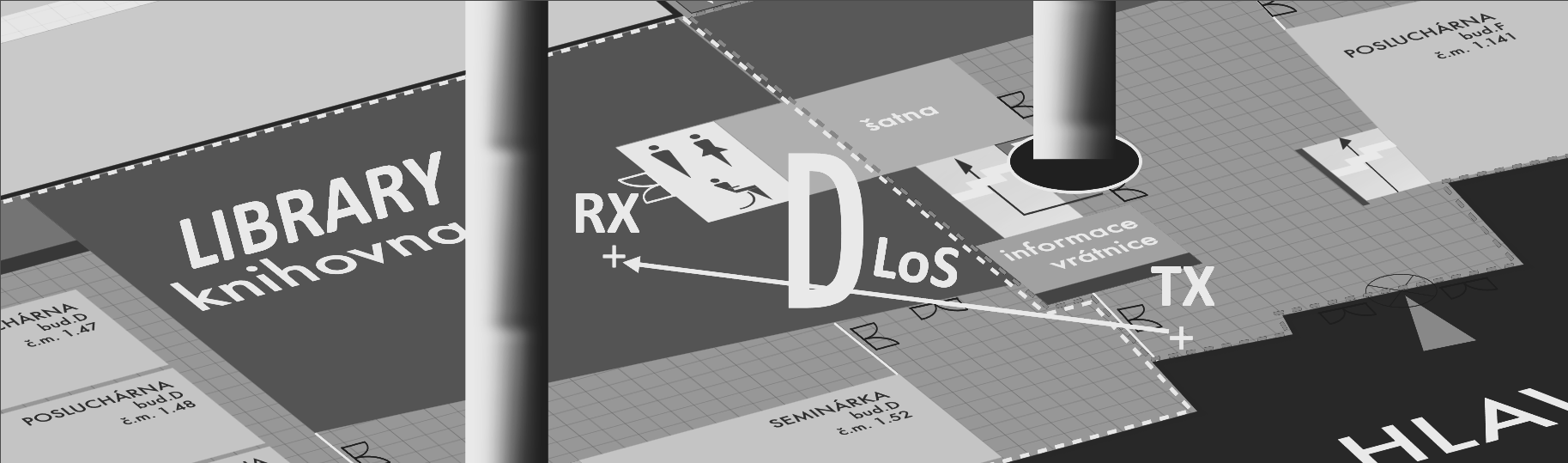}
    \caption{Plan of the measured scenario - floor plan}
    \label{fig:pudorys}
\end{figure}

The measurement campaign has been focused on the statistical channel description in a dynamic environment and has been performed at \ac{FEEC}, \ac{BUT}, Technicka 12, Brno, Czech Republic. \ac{TX} and \ac{RX} were in static position and environment was changing in time- pedestrians were present, the glass door was being opened and closed, causing \ac{MPC}. The \ac{TX} position was next to the main entrance to the building and \ac{RX} was placed in the faculty library with antennna orientation to the \ac{TX}. An open-ended waveguide antenna was used at the \ac{TX}. This setup demonstrated a base station with a wide beam antenna designed to cover a broader area. At the receiver, a horn antenna was employed to improve signal strength and to mitigate \ac{MPC}. Locations of \ac{TX} and \ac{RX} in the building are shown in a floor plan in Fig.~\ref{fig:pudorys} with \ac{LOS} highlighted.

The measurement campaign was focused on two frequency bands, 60\,GHz and 80\,GHz, respectively. As was mentioned in Chapter \ref{introduction}, \ac{AGV} could be used to distribute books from library or other equipment across institute in building. To ensure reliable communication and safe operation, it is necessary to understand the behavior of the \ac{MMW} propagation. \ac{TX} antenna was positioned in 1.4\,m and height of \ac{RX} antenna was variable in two position - 1.4\,m and 0.8\,m, respectively. The goal was to compare the standard operating height for a mobile phone in the hand or in a breast pocket and the height of the antenna on an \ac{AGV} for near ground communication. We captured a twenty-minute recording of each scenario to enable statistical evaluation of long-term channel measurements. One scenario at 60\,GHz, the RX antenna in a low position, was measured twice at different times to eliminate randomness and allow comparison of an identical scenario.

\section{Measurement setup}\label{section_MS}
\begin{figure*}[htbp]
    \centering
    \includegraphics[width=1\textwidth]{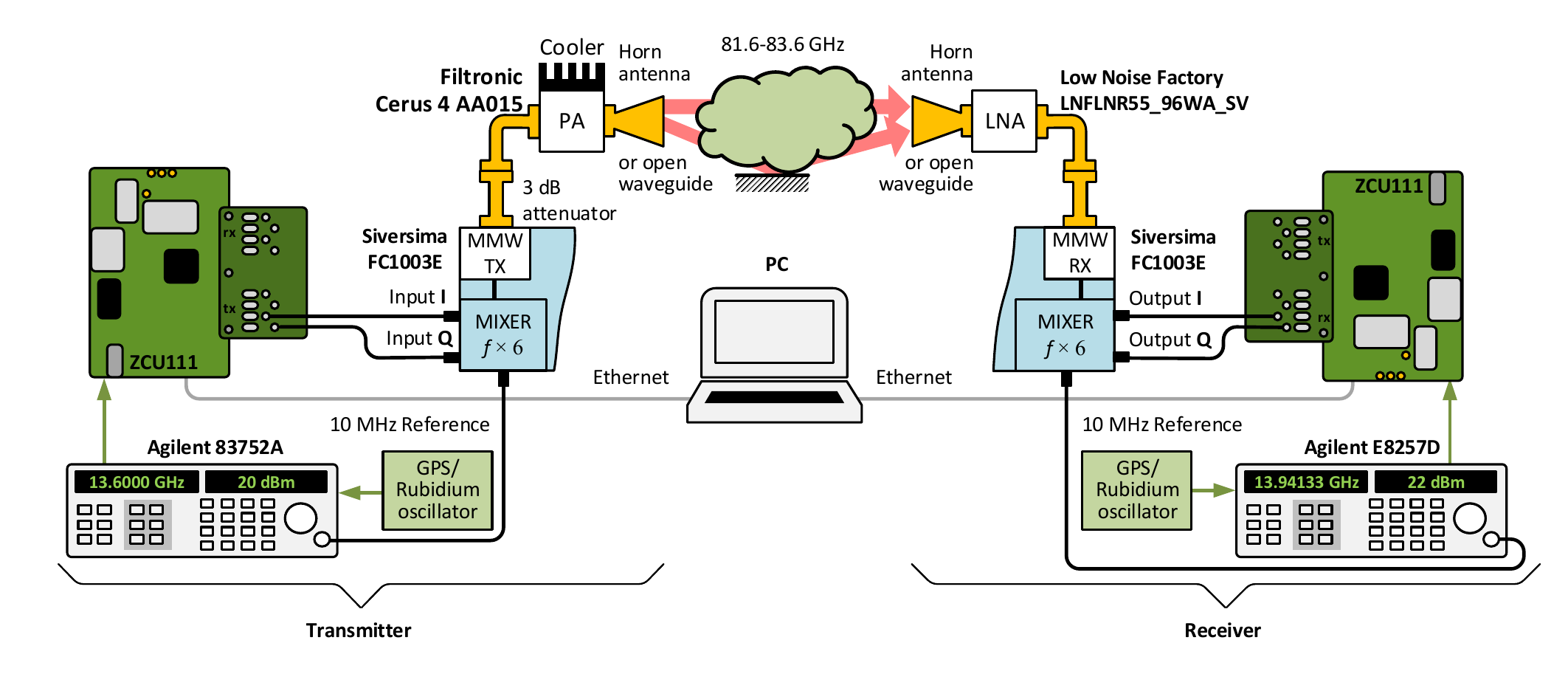}
    \caption{Measurement system schematic}
    \label{fig:meas_schema}
\end{figure*}

The schematic representation of the measurement setup is shown in Fig.~\ref{fig:meas_schema}. The board ZCU111 of Xilinx Zynq UltraScale+ RFSoC is used as a transmit baseband subsystem. The I and Q components of an intermediate frequency signal are produced by fast DACs clocked at 6.144\,GSPS. The excitation signal was a \ac{FMCW} with ramp-up and ramp-down to provide the flattest feasible spectrum.

This decision was primarily motivated by the fact that the \ac{FMCW} signal exhibits remarkable robustness in the presence of system non-linearities. With a bandwidth $B=2.048\,\mathrm{GHz}$ and duration $T=8\,\mu s$ , the \ac{FMCW} allows for fast measurements up to $f_{\mathrm{meas}}=\frac{1}{T}=125\,\mathrm{kHz}$ (measurements per second) while keeping an acceptable \ac{SNR}. Averaging can be used to further increase the \ac{SNR} in static scenario.

The Sivers IMA FC1005V/00 (60\,GHz) or FC1003E/03 (80\,GHz) up/down converter raises the signal to the required millimeter wave frequency. For the upconversion, a frequency stable, low phase noise local oscillator signal is supplied by the Agilent 83752A generator. The QuinStar QPW-50662330-C1 (60\,GHz) or Filtronic Cerus 4 AA015 (80\,GHz) power amplifier boosts the RF signal power, which is then transmitted using open-ended waveguide antenna. 

\begin{figure}[htbp]
    \centering
    \includegraphics[width=0.49\textwidth]{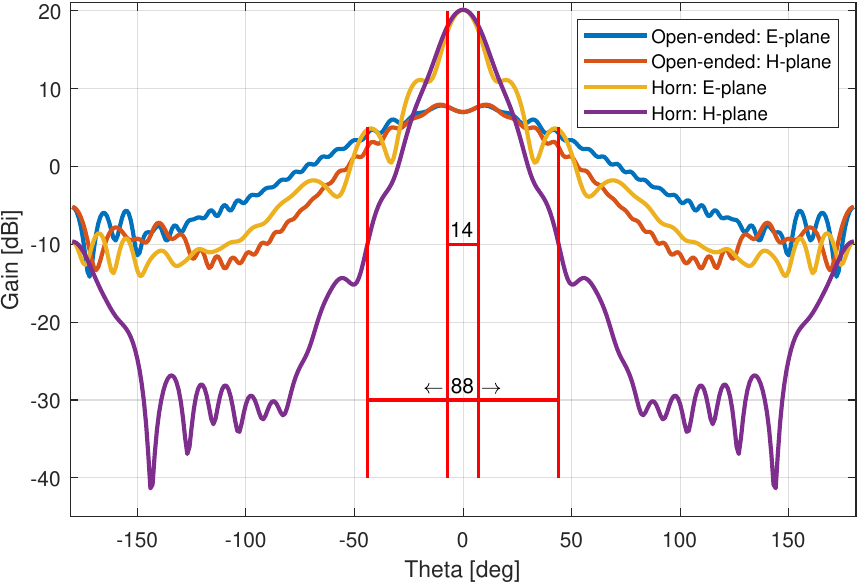}
    \caption{Simulated radiation pattern of \ac{OWGA} and horn antenna at 60\,GHz}
    \label{fig:rad_pat_60}
\end{figure}

\begin{figure}[htbp]
    \centering
    \includegraphics[width=0.5\textwidth]{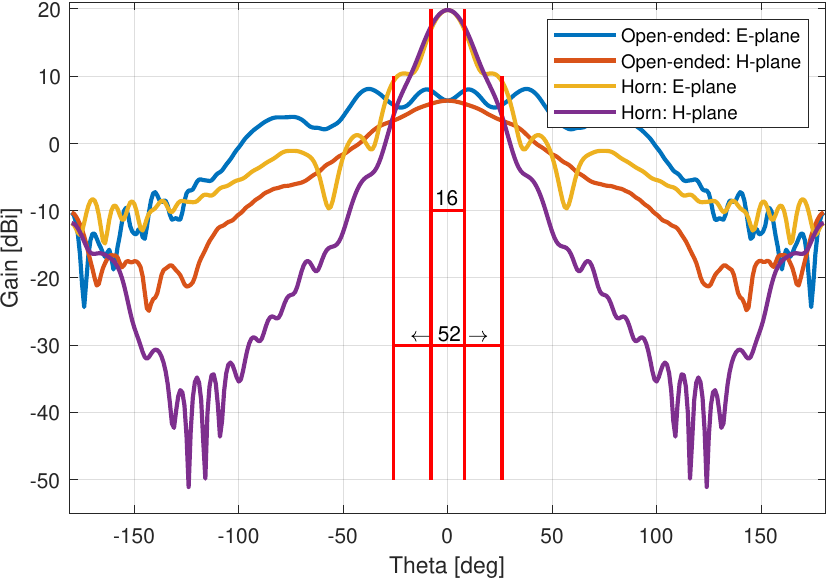}
    \caption{Simulated radiation pattern of \ac{OWGA} and horn antenna at 82\,GHz}
    \label{fig:rad_pat_80}
\end{figure}

The radiation patterns for the 60\,GHz \ac{OWGA} and horn antenna are shown in~Fig.~\ref{fig:rad_pat_60}. 
Fig.~\ref{fig:rad_pat_80} shows the radiation patterns for the 80\,GHz \ac{OWGA} and horn antenna. More information for both bands is provided in Tab. \ref{tab:ant_parameters60} and Tab. \ref{tab:ant_parameters80}.

\begingroup
\setlength{\tabcolsep}{20pt} 
\begin{table}[h]
    \centering
    \caption{Parameters of \ac{OWGA} and horn antenna at 60\,GHz}
    \begin{tabular}{l||c|c}
        &\ac{OWGA}&Horn\\
     \hline
      \hline
     Gain [dBi]&7& 20\\
     \hline
     HPBW E-plane [°]&88&14\\

    \end{tabular}
    
    \label{tab:ant_parameters60}
\end{table}
\endgroup
\begingroup
\setlength{\tabcolsep}{20pt} 
\begin{table}[h]
    \centering
    \caption{Parameters of \ac{OWGA} and horn antenna at 82\,GHz}
    \begin{tabular}{l||c|c}
        &\ac{OWGA}&Horn\\
     \hline
      \hline
     Gain [dBi]&6.4& 20\\
     \hline
     HPBW E-plane [°]&52&16\\

    \end{tabular}
    
    \label{tab:ant_parameters80}
\end{table}
\endgroup

A processing chain that is similar to the transmitting side of the measuring setup receives the signal after it has traveled through the measured environment. The Quinstar QLW-50754530-I2 (60\,GHz) or Low noise factory LNF-LNR55\_96WA\_SV (80\,GHz) low noise amplifier amplifies the signal after it is received by the horn antenna. Using a local oscillator signal produced by the Agilent E8257D generator, the Sivers IMA FC1003V/01 (60\,GHz) or FC1003E/02 (80\,GHz) up/down converter performs the downconversion. After being sampled at an intermediate frequency in the form of its I and Q components, the signal is recorded to an SSD for additional processing by using the fast ADCs (clocked at 4.096\,GSPS) of another Xilinx Zynq UltraScale+ RFSoC ZCU111 board. More information about the testbed and calibration process can be found in \cite{access_radek}.

\subsection{Principle of long-term data recording}\label{section_MS_LTM}
\begin{figure}[htbp]
    \centering
    \includegraphics[width=0.5\textwidth]{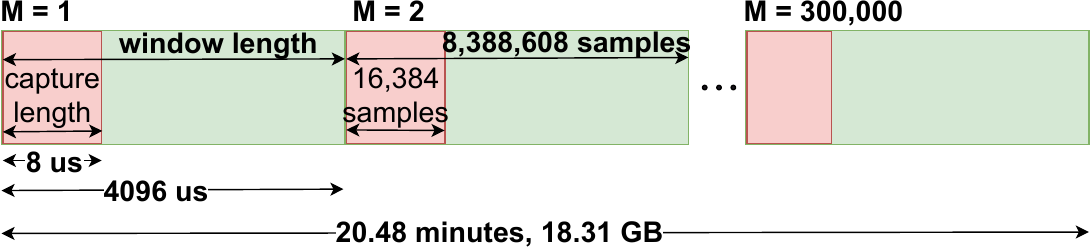}
    \caption{Graphic representation of long-term data recording}
    \label{fig:long_term_principle}
\end{figure}

The measurement setup, based on the Xilinx Zynq UltraScale+ RFSoC ZCU111 board as mentioned above, enables long-term data recording. In each clock cycle, 256\,bits (32\,bytes) of data are stored, with each sample comprising 4\,bytes (16\,bits I, 16\,bits Q). This allows for the storage of 8 samples per cycle. Given the large volume of data generated, continuous recording for tens of minutes is not feasible. The method for long-term data recording is illustrated in Fig.~\ref{fig:long_term_principle}. The window length is set to 8,388,608\,samples, corresponding to 4096\,µs. From each window, the first 16,384 samples, equivalent to 8 µs, are retained, while the remaining data is discarded. To store 20\,minutes of data, $M =$ 300,000 windows are recorded, corresponding to 1,228.8\,s and requiring 18.31\,GB of storage.


\section{Statistics of indoor channel}\label{section_statistic}
This section presents a statistical analysis of long-term measured data in the \mbox{60\,GHz and 80\,GHz} frequency bands, along with the extraction of parameters describing the channel.

We used our time domain channel sounder to estimate the \ac{CIR} \cite{Hlawatsch2011}
\begin{equation}
\label{rce:CIR}
h_m(t, \tau)=\sum_{n=1}^{N(t)} \alpha_n(t)\mathrm{e}^{j2\pi f_D t} \delta(\tau-\tau_n(t)),
\end{equation}

where $m$ is a measurement index and $N$ is the number of propagation paths. The variables $\alpha_n(t)$ and $\tau_n(t)$ corresponds to the amplitude and delay of the $n$-th propagation path while $\delta$ is the Dirac impulse and $f_D$ is the Doppler frequency.

\begin{figure}[htbp]
    \centering
    \includegraphics[width=0.48\textwidth]{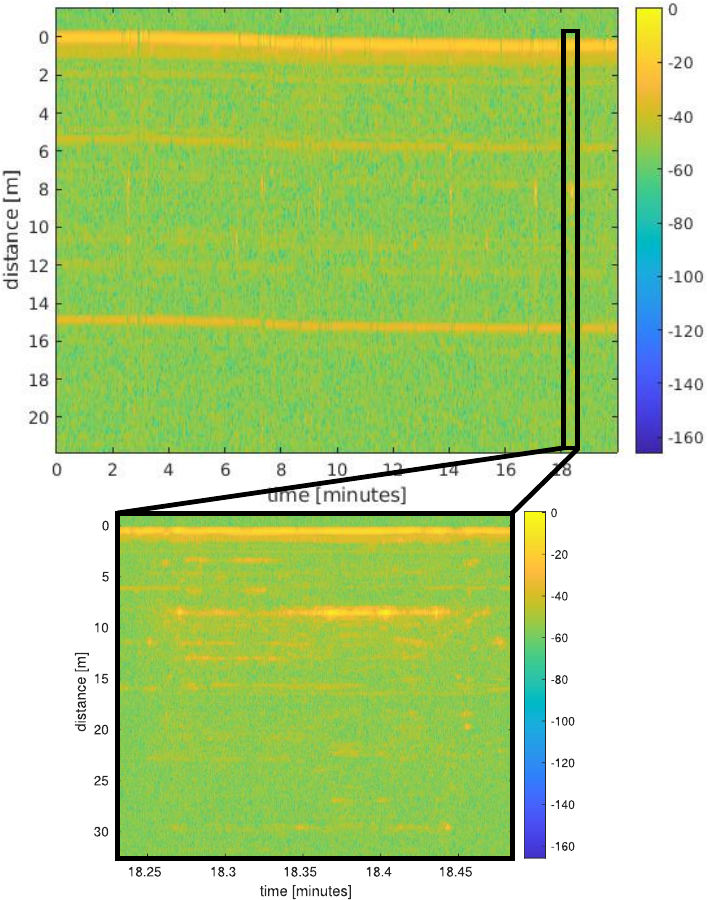}
    \caption{CIR of long-term measurement at 80\,GHz, \ac{RX} antenna at high position}
    \label{fig:CIR_scenario80_high}
\end{figure}

\begin{figure}[htbp]
    \centering
    \includegraphics[width=0.5\textwidth]{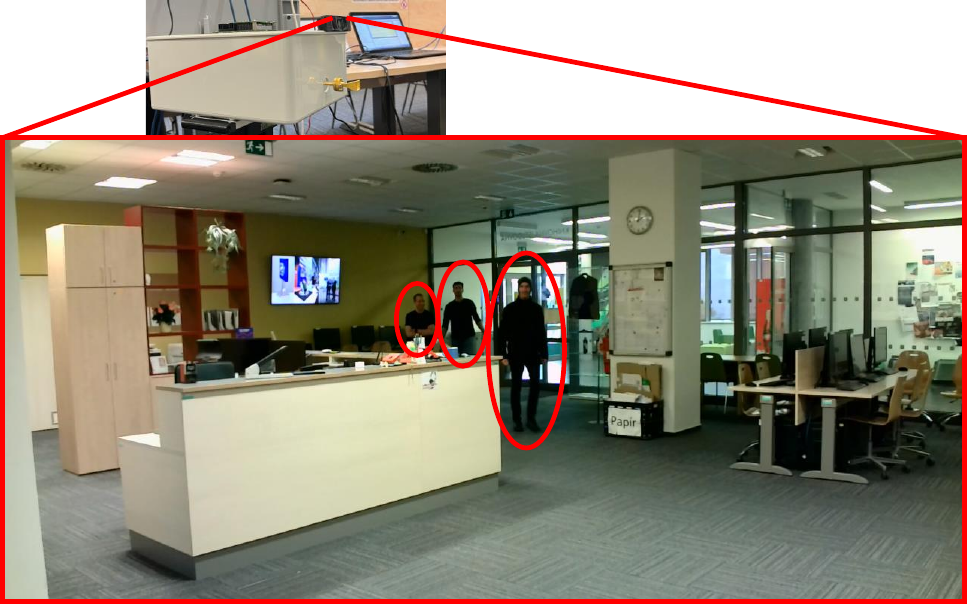}
    \caption{Video frame from the measurement campaign corresponding to the highlighted \ac{MPC} in the \ac{CIR} shown in Fig.~\ref{fig:CIR_scenario80_high}}
    \label{fig:CIR_video}
\end{figure}

Temporal variations in the CIR are illustrated in Fig.~\ref{fig:CIR_scenario80_high}. Observable deviations are attributed to environmental factors such as human movement and the opening or closing of doors, which occasionally obstructed the direct signal path. In addition to capturing the CIR as I and Q samples, video footage was recorded from the perspective of the RX antenna to facilitate the analysis of the specific causes of \ac{MPC} deviations. A video frame corresponding to the specific \ac{MPC} of CIR highlighted in Fig.~\ref{fig:CIR_scenario80_high} is presented in Fig.~\ref{fig:CIR_video}.


\subsection{RMS delay spread}\label{section_statistic_RMS}
The \ac{RMS} delay spread is calculated from \ac{PDP} according to \cite{molish_RMS_delay_spread} : 

\begin{equation}
\label{rce:RMS_delay_spread}
\sigma_{\tau, n} = \sqrt{\frac{\sum_{L}^{i=1} P(\tau_i,n) \tau_i^2}{\sum_{L}^{i=1} P(\tau_i,n)}-\frac{(\sum_{L}^{i=1} P(\tau_i,n) \tau_i)^2}{(\sum_{L}^{i=1} P(\tau_i,n))^2}},
\end{equation}
where $P(\tau_i,n) = E\{|h(\tau_i,n)|^2\}$ are the \ac{PDP} taps at the delay~$\tau_i$, $L$ is the number of taps and $h(\tau_i,n)$ denotes the complex channel impulse response at the delay~$\tau_i$ for the $n$-th measurement.

\begin{figure}[htbp]
    \centering
    \includegraphics[width=0.48\textwidth]{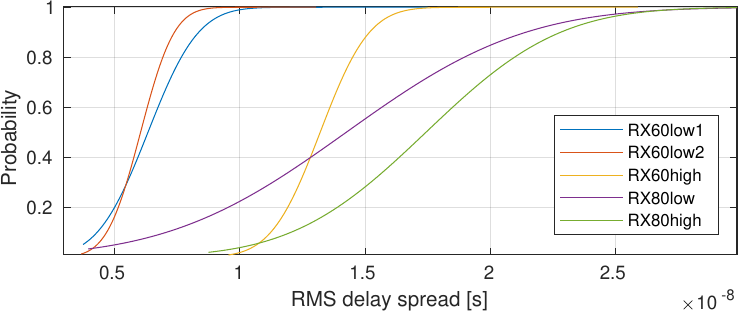}
    \caption{Cumulative distribution function of RMS Delay Spread}
    \label{fig:CDF_RMSDS}
\end{figure}

We recorded 20 minutes of data for each scenario, allowing us to compute how the \ac{RMS} delay spread changes over time. The \ac{CDF} of the \ac{RMS} delay spread is depicted in Fig.~\ref{fig:CDF_RMSDS}. It is visible that the shape of the \ac{CDF} is consistent across the frequency bands (blue, red vs. yellow curve (60\,GHz) and purple vs. green (80\,GHz) ); however, there is a shift along the x-axis depending on the height of the \ac{RX} antenna. In one case, we performed the measurement twice for the same configuration (60\,GHz and antenna at a low position) to eliminate randomness. It was confirmed that the \ac{CDF} for both measurements is almost the same.


The RMS delay spread is higher at 80\,GHz due to the shorter wavelength, which increases the likelihood of reflections from smaller objects. Additionally, significant differences are observed based on antenna height. Antennas positioned at higher elevations capture more \ac{MPC}, leading to an increased RMS delay spread compared to antennas at lower positions. This trend is evident in both frequency bands.


\subsection{Rician $K$-factor}\label{section_statistic_Kfac}
The Rician $K$-factor is given by
\begin{equation}
\label{rce:K_fac}
K = 10 \log_{10}\left(\frac{r^2}{2\sigma^2}\right),
\end{equation}
where $r^2$ represents the power of the \ac{LOS} component and $2\sigma^2$ denotes the variance of the \ac{MPC} \cite{K_factor5}.

\begin{figure}[htbp]
    \centering
    \includegraphics[width=0.48\textwidth]{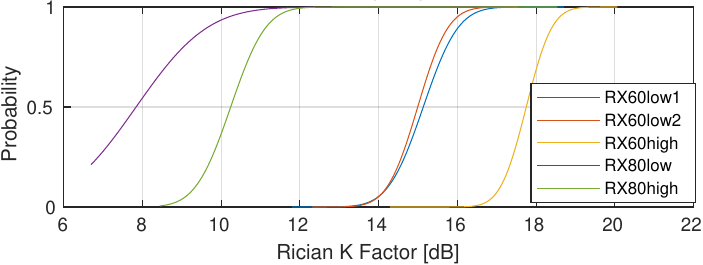}
    \caption{Cumulative distribution function of Rician $K$-factor}
    \label{fig:CDF_Kfac}
\end{figure}

The \ac{CDF} of the Rician $K$-factor is depicted in Fig.~\ref{fig:CDF_Kfac}. The distribution of the graph curves corresponds to the assumption according to the results from \ac{RMS} delay spread. 

The highest $K$-factor is observed at 60\,GHz with the antenna positioned at a higher elevation. This is attributed to the longer wavelength at this lower frequency, which results in fewer \ac{MPC} and a stronger line-of-sight (LOS) component relative to the \ac{MPC}. The $K$-factor decreases by approximately 3\,dB when the antenna is positioned lower. In the E-band, the increased reflections from various objects result in a lower $K$-factor. Similarly, antenna height plays a significant role: a higher antenna position corresponds to a higher $K$-factor, and a lower antenna position corresponds to a lower $K$-factor, respectively.




\section{Conclusion}\label{section_C}
The primary contributions of this paper include an analysis of a dynamic channel in indoor scenarios, where the positioning of the antennas is used to simulate differences between \ac{AGV} and pedestrian communication environments. Additionally, we examine the impact of moving people and objects in the vicinity of the \ac{TX} and \ac{RX} antennas, and their influence on radio channel properties. Our analysis primarily focuses on the evaluation of the \ac{RMS} delay spread and the Rician $K$-factor, revealing a significant influence of antenna height and radio signal frequency. The \ac{RMS} delay spread is higher in the E-band by approximately 5–10\,ns, and it also increases for higher antenna positions at the same frequency band. In line with these findings, the $K$-factor is higher at 60\,GHz and decreases by approximately 3\,dB when the antenna is positioned lower. These insights into the behavior of high-frequency dynamic channels in environments relevant to \ac{AGV} operations contribute to the development of more robust and reliable communication systems for such applications.


\section*{Acknowledgment}
The research described in this paper was financed by the Czech Science Foundation, Project No. 23-04304L, Multi-band prediction of millimeter-wave propagation effects for dynamic and fixed scenarios in rugged time varying environments and by the Internal Grant Agency of the Brno University of Technology under project no. FEKT-S-23-8191. The work of A. Chandra is supported by the Chips-to-Startup (C2S) program no. EE-9/2/2021-R\&D-E from MeitY, GoI. The work of Jan M. Kelner, Jarosław Wojtuń and Cezary H. Ziółkowski was funded by the National Science Centre, Poland, under the OPUS-22 (LAP) call in the Weave program, as part of research project no. 2021/43/I/ST7/03294, acronym ‘MubaMilWave’.

\bibliographystyle{IEEEtran}
{\footnotesize
\bibliography{IEEEabrv,biblio_yousef.bib}
}


\end{document}